\documentclass[letter]{aa}
\usepackage{graphicx}
\usepackage{nicefrac}
\usepackage{multicol}
\usepackage{natbib,twoopt}
\usepackage[breaklinks=true]{hyperref}
\usepackage[varg]{txfonts}
 \bibpunct{(}{)}{;}{a}{}{,}   
 \newcommandtwoopt{\citeads}[3][][]{\href{http://adsabs.harvard.edu/abs/#3}%
                                        {\citealp[#1][#2]{#3}}}
 \newcommandtwoopt{\citepads}[3][][]{\href{http://adsabs.harvard.edu/abs/#3}%
                                        {\citep[#1][#2]{#3}}}
 \newcommandtwoopt{\citetads}[3][][]{\href{http://adsabs.harvard.edu/abs/#3}%
                                        {\citet[#1][#2]{#3}}}
 \newcommandtwoopt{\citealtads}[3][][]{\href{http://adsabs.harvard.edu/abs/#3}%
                                        {\citealt[#1][#2]{#3}}}
 \newcommandtwoopt{\citeyearads}[3][][]%
   {\href{http://adsabs.harvard.edu/abs/#3}{\citeyear[#1][#2]{#3}}}
\begin{document}

\title{Planet-induced disk structures: A comparison between (sub)mm and infrared radiation}

\author{Jan Philipp Ruge\inst{\ref{inst1},}\thanks{ruge@astrophysik.uni-kiel.de} \and Sebastian Wolf\inst{\ref{inst1}} \and Ana L. Uribe\inst{\ref{inst2},\ref{inst3}}, Hubert H. Klahr\inst{\ref{inst2}}}
\institute{Universit\"at zu Kiel, Institut f\"ur Theoretische und Astrophysik, Leibnitzstr. 15, 24098 Kiel, Germany \label{inst1} 
\and
Max-Planck-Institut f\"ur Astronomie, K\"onigstuhl 17, 69117 Heidelberg, Germany \label{inst2}
\and
University of Chicago, The Department of Astronomy and Astrophysics, 5640 S. Ellis Ave, Chicago, IL 60637, USA \label{inst3}}

\date{Received 06. November 2013 / Accepted 05. November 2014 }

 \abstract{Young giant planets, which are embedded in a circumstellar disk, will significantly perturb the disk density distribution. This effect can potentially be used as an indirect tracer for planets.}{We investigate the feasibility of observing planet-induced gaps in circumstellar disks in scattered light.}{We perform 3D hydrodynamical disk simulations combined with subsequent radiative transfer calculations in scattered light for different star, disk, and planet configurations. The results are compared to those of a corresponding study for the (sub)mm thermal re-emission.}{The feasibility of detecting planet-induced gaps in scattered light is mainly influenced by the optical depth of the disk and therefore by the disk size and mass. Planet-induced gaps are in general only detectable if the photosphere of the disks is sufficiently disturbed. Within the limitations given by the parameter space here considered, we find that gap detection is possible in the case of disks with masses below $\sim 10^{-4\dots-3} \, \rm M_\odot$. Compared to the disk mass that marks the lower Atacama Large (Sub)Millimeter Array (ALMA) detection limit for the thermal radiation re-emitted by the disk, it is possible to detect the same gap both in re-emission and scattered light only in a narrow range of disk masses around $\sim 10^{-4} \, \rm M_\odot$, corresponding to $16\%$ of cases considered in our study.}{}


\titlerunning{Detecting planet-induced disk structures}
\authorrunning{Ruge et al.}
\maketitle

\section{Introduction}
Observations obtained with high angular resolution, high contrast near-infrared observation techniques show strong evidence of gap structures and young giant planets in face-on circumstellar disks (e.g., \citealtads{2013ApJ...766L...1Q,2013ApJ...766L...2Q}). However, the interpretation of these scattered light images as indicators of planet-disk interaction is quite uncertain because shadowing and wavelength-dependent optical depth effects may result in similar structures in the observed images \citepads[e.g.,][]{2011A&A...527A..27S}. Additionally, nonaxisymmetric structures in the disk density distribution, which are not induced by planet-disk interaction, have been predicted (e.g. \citealtads{2002ApJ...578L..79W,2010MNRAS.407..181C,2013A&A...549A..97R}). Therefore, it is necessary to perform complementary and independent observations of the potential planet-forming regions in circumstellar disks to disentangle the individual underlying scenarios \citepads{2008PhST..130a4025W}.
A promising means to accomplish this is offered by the thermal disk emission in the (sub)mm wavelength regime, where the density profiles of disks in face-on orientations and, thus, planet-induced gaps can be traced directly (e.g., \citealtads{2005ApJ...619.1114W}, \citealtads{2012A&A...547A..58G,2013A&A...549A..97R}). Beginning about two years ago, the Atacama Large (Sub)Millimeter Array (henceforth ALMA) has allowed one to perform first of these observations (e.g.,\citealtads{2013Sci...340.1199V,2013ApJ...775...30I,2014ApJ...783L..13P}).\par
In the following, we present our investigations of the near-infrared scattered light detectability of planet-induced gaps in circumstellar disks. Earlier studies for this wavelength range have been performed by \citetads{2006ApJ...637L.125V}, \citetads{2008PhST..130a4025W}, and \citetads{2009ApJ...700..820J}, but were limited to the use of 2D hydrodynamical (henceforth HD) calculations of the disk structure with an assumed Gaussian vertical profile. In contrast, our approach is based on 3D HD calculations and follow-up radiative transfer calculations. The resulting scattered light images are compared to the outcome of a corresponding study concerning thermal disk re-emission \citepads{2013A&A...549A..97R}.\\
We consider disk masses between $10^{-6} \, \rm M_\odot$ and $10^{-2} \, \rm M_\odot$, disk outer radii ranging from $90 \, \rm AU$ to $225 \, \rm AU$, and two types of central stars. The distance of the object is assumed to be $140\, \rm pc$ and the observability of the planet-induced gap is investigated for various disk inclinations, from face- to edge-on.
  \begin{figure}[htbp]
      \centering
      \resizebox{0.85\hsize}{!}{\includegraphics{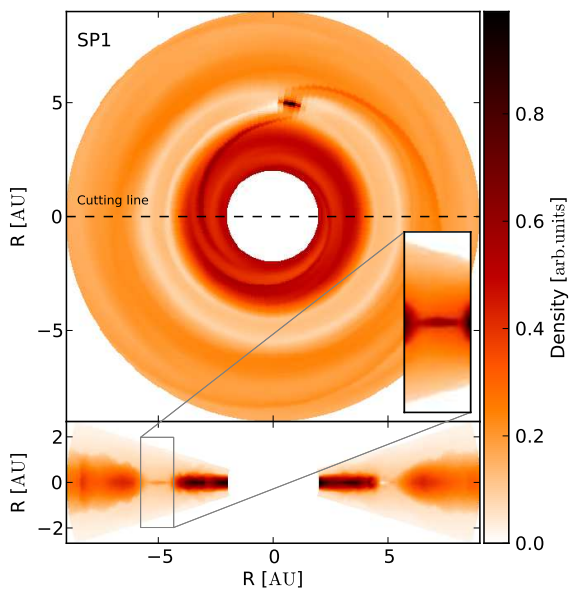}}
  \caption{Density distribution of the disk model \textbf{SP1} in the midplane of the disk and along a vertical cut through the disk (indicated by the dashed line in the midplane density plot) in arbitrary units in logarithmic scale. The vertical density distribution of the gap is shown as an inset with a different color-scale.}
  \label{fig:sp1_den}
  \end{figure}
\section{Modeling techniques}
\label{sec:sim}
\paragraph{Hydrodynamical simulations:}
\label{sec:hydro}
We focus on ideal isothermal 3D HD disk simulations with variations of scale height $\nicefrac{H}{r} \in \{0.03,0.07\}$ and $\alpha$-viscosity $\alpha \in \{0, 10^{-3}, 10^{-2}\}$. The disks are perturbed by a planet with a planet-to-star mass ratio $q$ of $10^{-3}$ or $2 \times 10^{-3}$. These parameters are essential for the properties of a planet-induced gap. The depth and width of a planet-induced gap in the disk density distribution increases both with increasing planet-to-star mass ratio and decreasing scale height and viscosity \citepads{1984ApJ...285..818P}. For example, Fig.~\ref{fig:sp1_den} shows the disk density profile along a vertical and horizontal cut for a disk model with $\alpha = 0$, $q= 10^{-3}$, and $\nicefrac{H}{R} = 0.07$, which is equal to model \textbf{SP1} in \citealtads{2013A&A...549A..97R}. It will be referred to as \textbf{SP1} here. All simulation techniques and assumptions used here are described in detail in \citetads{2013A&A...549A..97R}.
\paragraph{Radiative transfer:}
The scattered light radiative transfer simulations are performed with the Monte Carlo-based 3D continuum radiative transfer code MC3D (\citealtads{1999A&A...349..839W,2003CoPhC.150...99W}). The dust consists of $62.5\, \%$ silicate and $37.5\, \%$ graphite (optical data from \citealtads{2001ApJ...548..296W}) with a dust density of $\rho_\text{dust} = 2.7\, \rm g\,cm^{-3}$. The dust grain size distribution follows a power law, $n(a) \propto a^{-2.5}$ for $a \in \left[0.005\, \mu \textup{m}, 100 \, \mu \textup{m}\right]$ \citepads{1969JGR....74.2531D}. The dust properties are identical to the case of \textbf{large dust} in \citetads{2013A&A...549A..97R}. The star and planet are both considered radiation sources in blackbody approximation.
\label{sec:setup}
\paragraph{Setup:}
The considered parameter space is a subset of that presented in \citetads{2013A&A...549A..97R}.
The HD models are scale free, which allows us to scale them in mass and size. In the unscaled case, the disks range from $2\, \rm AU$ to $9\, \rm AU$ with the planet at $5\, \rm AU$. The disk size is adapted by scaling the disk in the radial direction with a factor $k \in \left\{10, 13, 16, 19, 22, 25\right\}$. The total disk mass is considered to be $M = 2.7 \times 10^{-6 \cdots -1} \, \rm M_\odot$, with a dust to gas mass ratio of $\nicefrac{1}{100}$. For the central star, we consider a T Tauri star with $T = 4000\, \rm K$, $L = 0.95 \, \rm L_\odot$, and $M = 0.5\, \rm M_\odot$, and a Herbig AE star with $T = 9500\, \rm K$, $L = 43 \, \rm L_\odot$, and $M = 2.5\, \rm M_\odot$.
As in \citetads{2005ApJ...619.1114W}, we assume a planet with $T = 1000 \, \rm K$ and $L = 10^{-4}\, \rm  L_\odot$. We calculate the synthetic scattered light images for the wavelengths $1.0\, \rm \mu m, 2.2\, \rm \mu m, 4.0\, \rm \mu m,$ and $5.0 \, \rm \mu m$, and inclinations of the object of $i \in \left\{5^\circ, 15^\circ, 30^\circ, 45^\circ, 60^\circ, 75^\circ, 90^\circ \right\}$. The thermal emission of the disk is neglected in the selected wavelength range. This assumption is verified by a small ratio between the thermal disk emission and scattered radiation. The maximum ratio, which is reached in the entire parameter space, is $0.12\%$ at a wavelength of $5.0\, \rm \mu m$.
\paragraph{Analysis:}
\label{sec:anaylsis}
   \begin{figure}[t]
	\centering
        \resizebox{0.95\hsize}{!}{\includegraphics{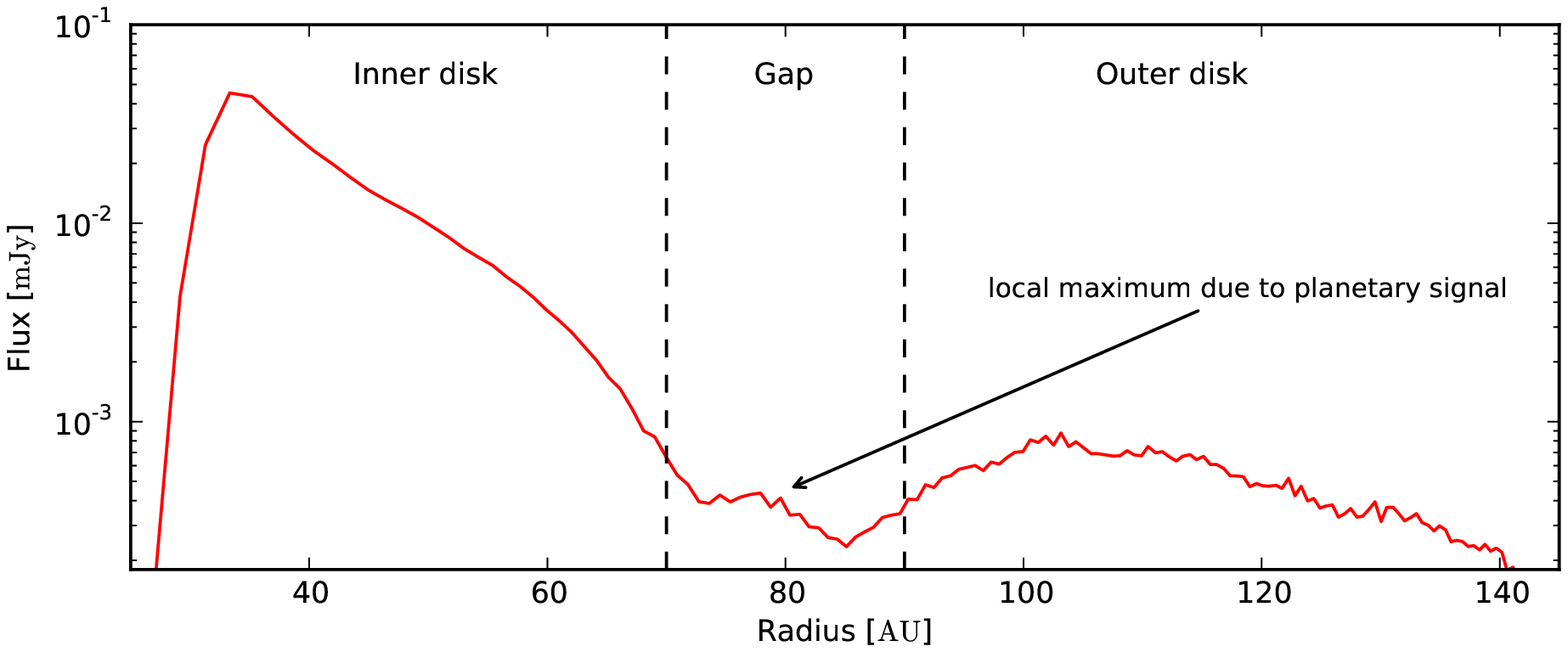}}
   \caption{Example of the brightness profile that is used in the assessment process for detecting gaps discussed in \S \, \ref{sec:anaylsis} in detail. The presence of a gap at $80\, \rm AU$ is indicated. The total disk mass is $2.7 \times 10^{-4} \, \rm M_\odot$ and the disk is oriented face-on. The underlying disk model is SP1. The star is a T Tauri star and the results are shown for a wavelength of $2.2\, \rm \mu m$.}
   \label{fig:bild}
   \end{figure}
The tidal interaction between a circumstellar disk and an embedded giant planet ($\approx 0.1\%$ of the stellar mass) will induce an axisymmetric gap in the disk density distribution along the entire orbit of the planet \citepads{1984ApJ...285..818P}. In particular, this azimuthal symmetry of the gap is interesting. To trace planet-induced gaps in ideal scattered light images, we calculate the azimuthally averaged and inclination-corrected brightness profile of the disk. For moderate disk inclinations ($i \leq 45^\circ$), this helps us to achieve a maximum of signal-to-noise. \par
For example, Fig. \ref{fig:bild} shows the brightness profile of a disk model with the planet at a semi-major axis of $80\, \rm AU$ in a face-on orientation. The minimum in the profile at roughly $80\, \rm AU$ is directly linked to the gap formed by the planet in the disk density distribution (see Fig. \ref{fig:sp1_den}). This minimum divides the profile into the signals of the inner and outer parts of the disk. In the inner part of the disk, the flux of the scattered light reaches its global maximum.\par
To detect a gap, we assess all brightness profiles in the following way. First, we search the characteristic flux minimum in the brightness profile $F_0$. After that, we search for the maximum flux $F_\text{max}$ at radii larger than the location of the minimum. Then, we define the contrast ratio $\zeta = \nicefrac{(F_\text{max} - F_0)}{F_\text{max}}$.
A gap is considered as detected in the brightness profile if $\zeta > 0.1$ and the location of the minimum is consistent (deviation smaller than the width of the gap) with the semi-major axis of the planet. Additionally, the difference between the maximum $F_\text{max}$ and the flux value at the orbital position of the planet $F_0$ has to be larger than three times the noise in the scattered light images due to the Monte Carlo nature of the calculations. In our simulations, the planet contributes a direct signal to the brightness profile, this more accurately approximates what one would observe.\par
For inclinations above $45^\circ$, we use the brightness distribution of a cut through the disk along the dashed line in Fig. \ref{fig:sp1_den}.
These brightness cuts are then assessed in the same way as for lower inclinations. Through the position of the cut, the planetary signal is not considered in this assessment.\par
In general, this method avoids the false identification of gaps, created by shadowing effects, as planet-induced gaps because we only consider gaps at the exact planetary position. Additionally, gaps created by shadowing effects will not show the strong disk mass dependency we present in the results section (\S \,  \ref{sec:results}) because for each case they are limited to a particular disk mass. 
The planetary signal is only discussed in course of the assessment of the brightness profiles. Further results dealing with the direct planetary signal are not presented yet.
\section{Results}
\label{sec:results}
   \begin{figure*}[htbp]
	\sidecaption
       \includegraphics[width = 12cm]{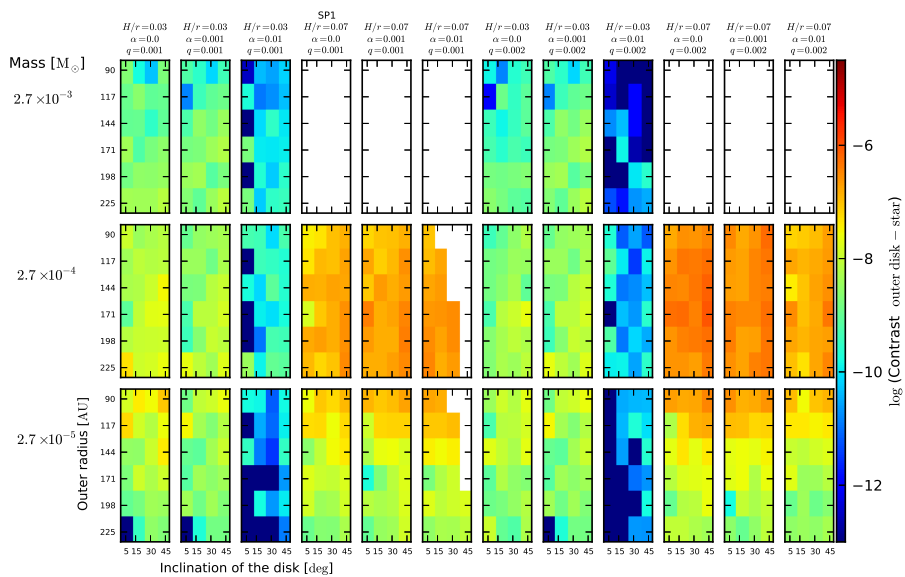}
  \caption{Exemplary extraction of the results of this study for the wavelength $\lambda = 2.2\, \rm \mu m$. If a combination of the parameters wavelength, disk size, and mass (focus on masses between $10^{-3}\, \rm M_\odot$ and $10^{-5}\, \rm M_\odot$) and inclination (focus on inclinations $\leq 45^\circ$) leads to a possible gap detection (see \S \, \ref{sec:anaylsis}), the contrast between outer disk and star is color-coded. The colorbar is presented in log scale. Each disk model, indicated by the parameters scale height ($\nicefrac{H}{r}$), viscosity ($\alpha$), and planet-to-star mass ratio ($q$), is presented in its own column. The fourth column presents the results for the disk model \textbf{SP1}, which is selected as an example in the text. The central object is a T Tauri star.}
  \label{fig:1um}
   \end{figure*}
   \begin{figure}[htbp]
	\centering
       \resizebox{0.9\hsize}{!}{\includegraphics{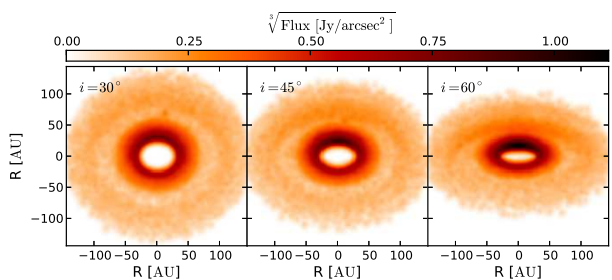}}
   \caption{Example of the synthetic object images of the disk simulation \textbf{SP1} at $2.2\, \rm \mu m$ for three inclinations listed top left in each plot. The disk has an outer radius of $144\, \rm AU$ and a total mass of $2.7 \times 10^{-4} \, \rm M_\odot$. The star is a Herbig Ae. The direct stellar radiation has been removed.}
   \label{fig:scat_1625}
   \end{figure}
   \begin{figure}[htbp]
	\centering
       \resizebox{0.9\hsize}{!}{\includegraphics{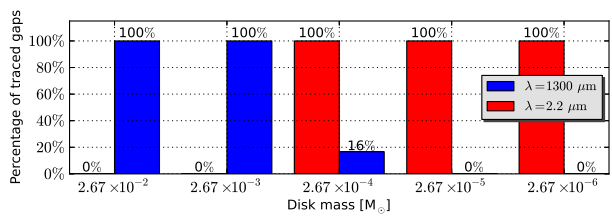}}
   \caption{Comparison of the number of traced gaps in case of $2.2\, \rm \mu m$ scattered light and the thermal re-emission case at $1300\, \rm \mu m$ taken from \citepads{2013A&A...549A..97R} for the disk model \textbf{SP1}.}
   \label{fig:histo}
   \end{figure}
Figure \ref{fig:1um} summarizes the outcome of our study for a wavelength of $2.2\,\rm \mu m$. If a combination of the parameters inclination, disk size, and mass, central star and wavelength lead to a detection of a planet-induced gap (in course of our definition in \S \, \ref{sec:anaylsis}) the corresponding contrast level between the outer disk ($F_\text{max}$) and star is plotted color-coded and in logarithmic scale into the figure. Figure \ref{fig:scat_1625} provides an example of the ideal scattered light images, which are used in this study. 
\paragraph{Disk mass and size:}
As the main result, we find that gaps in our setup are only detectable for total disk masses below $2.7 \times 10^{-4} \, \rm M_\odot$ in the case of a scale height of $\nicefrac{H}{r} = 0.07$ and $2.7 \times 10^{-3} \, \rm M_\odot$ in the case of a scale height of $\nicefrac{H}{r} = 0.03$. For the disk model \textbf{SP1} ($\nicefrac{H}{r} = 0.07$), this is also the lower limit for detections of gaps by ALMA in the (sub)mm wavelength range, found by \citetads{2013A&A...549A..97R}. The histogram in Fig. \ref{fig:histo} shows for different disk masses the distribution of traceable gaps both for the scattered light at $2.2\, \rm \mu m$ (blue) and for the thermal disk emission at $1300\, \rm \mu m$ with ALMA (red).
It shows, that it is in the majority of cases not possible to detect a planet-induced gap both in scattered light and in the (sub)mm disk re-emission with ALMA. 
This suggests that a mutual confirmation of observing results is feasible only in a very narrow range of disk masses ($16\%$ of cases for $2.7 \times 10^{-4}\, \rm M_\odot$, see Fig. \ref{fig:histo}).\par
   \begin{figure*}[t!]
    \sidecaption
   \includegraphics[width=12cm]{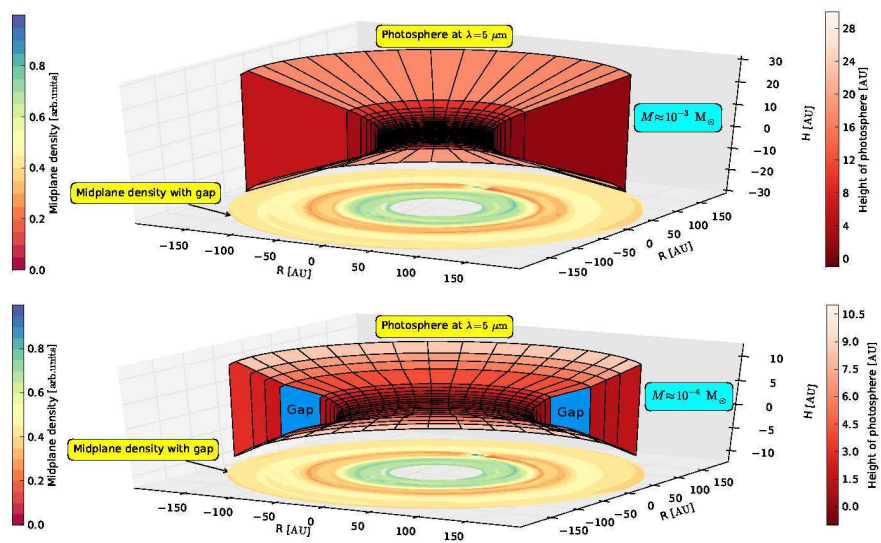}
   \caption{Photosphere of the selected disk model \textbf{SP1} ($R_\text{in} = 38 \, \rm AU$, $R_\text{out} = 198 \, \rm AU$, $\nicefrac{H}{r} = 0.07$) at $\lambda = 5.0 \, \rm \mu m$ and with two different disk masses. The gap is detectable in the scattered light if the gap is traced by a variation of the gradient in the radial direction of the vertically integrated optical depth. This is only the case for a disk mass equal or below $2.7 \times 10^{-4} \, \rm M_\odot$, as shown on the bottom.}
   \label{fig:tau1}
   \end{figure*}
The reason for this behavior is the optical depth structure of the disk given by the density distribution and the wavelength-dependent optical properties of the dust. The scattered light in the near- and mid-infrared range traces the photosphere of the disk corresponding to an optical depth equal to one for the selected wavelength. The gap is only traced in the final image if the photosphere shows evidence of a gap. Thereby, a gap is visible if the local gradient in the radial direction of the photosphere changes significantly. In our 3D HD disk simulations (see Fig. \ref{fig:sp1_den}) the density above the midplane in the zone of the gap is still large enough to create a significant contribution to the optical depth. For total disk masses above  $2.7 \times 10^{-4} \, \rm M_\odot$ ($\nicefrac{H}{r} = 0.07$) and $2.7 \times 10^{-3} \, \rm M_\odot$ ($\nicefrac{H}{r} = 0.03$) this is sufficient to completely mask the gap. Figure \ref{fig:tau1} shows the photospheres of a disk for the disk masses $2.7 \times 10^{-3}\, \rm M_\odot$ and $2.7 \times 10^{-4}\, \rm M_\odot$ for a selected rescaled version of the model \textbf{SP1} ($R_\text{in} = 38 \, \rm AU$, $R_\text{out} = 198 \, \rm AU$, $\nicefrac{H}{r} = 0.07$) at a wavelength of  $\lambda = 5.0 \, \rm \mu m$. The gap is invisible for a disk mass of $2.7 \times 10^{-3}\, \rm M_\odot$, while for $2.7 \times 10^{-4}\, \rm M_\odot$ the gap is identified by a variation of the gradient of the profile. Although the density distribution of the disk is strongly perturbed by the planet, both photospheres shown in Fig. \ref{fig:tau1} are nearly rotationally symmetric. This additionally masks the presence of a massive substellar object in the disk.\par
The turbulence through viscosity directly influences the depth of the planet-induced gap. For this reason, the number of detectable gaps is reduced for the disk model with $\nicefrac{H}{r} = 0.07$, $\alpha = 0.01$ and $q=0.001$ (see Fig. \ref{fig:1um}). For lower viscosities and higher planet-to-star mass ratios, turbulence only slightly influences the results, however, additional variations of the photosphere can be expected once 3D HD calculations consider complete thermodynamics of the disks.\par
For a static scale height, the number of traced gaps does not differ for either star in our setup. This is due to the linear scaling between scattered and incident radiation flux. However, the stellar component has a strong impact on disk temperature distribution. The scale height of the disk again depends on the temperature distribution. The stellar component, therefore, affects the detectability of gaps in scattered light by its influence on the scale height.\par
Since the optical depth decreases with increasing wavelength, the feasibility of detecting gaps increases with wavelength. Tests using ideal $20\, \rm \mu m$ images (including both scattered and re-emitted radiation) showed that it is possible to move the upper limit in disk mass (up to a factor of 10). Additionally, depending on the wavelength, we notice that the planet can be identified as an embedded source in the disk, although a gap detection is not feasible.\par
 
\section{Conclusion}
\label{sec:conclusions}
We present a feasibility study to identify planet-induced gaps in circumstellar disks through high-angular resolution observations at near-infrared wavelengths  ($1.0\, \rm \mu m$ to $5.0\, \rm \mu m$). We compare our results to those obtained in the (sub)mm range \citepads{2013A&A...549A..97R}. Our main results are:
\begin{itemize}
 \item To determine the correct shape of the wavelength-dependent photosphere of the disks, a 3D disk model including the vertical disk density structure is indispensable because simpler approaches (e.g., Gaussian flaring in vertical direction) can overestimate the detectability of planet-induced gaps.
 \item Based on our disk models, within a mass range from $10^{-1}\, \rm M_\odot$ to $10^{-6}\, \rm M_\odot$, we conclude that planet-induced gaps can only be traced in scattered light for total disk masses below the upper limit of $\approx 10^{-4} \, \rm M_\odot$ in the case of a scale height of $\nicefrac{H}{r} = 0.07$ and $\approx 10^{-3} \, \rm M_\odot$ in the case of a scale height of $\nicefrac{H}{r} = 0.03$. A decreasing scale height of the disk can increase the upper mass limit (in our setup up to a factor of $10$). The upper limit for scattered light was found by \citetads{2013A&A...549A..97R} to be the lower limit for gap detections in the (sub)mm wavelength range with ALMA for our disk model \textbf{SP1}. Because of that, the observation of gaps could possibly be mutually exclusive in either scattered or (sub)mm re-emitted light with today's instruments. 
 \item As the next important step, the continuation of our study has to focus on the detailed structure of the optically thin upper layers of the disk because the electromagnetic and particle radiation of the stellar component is expected to have an affect of major interest on them.
\end{itemize}

\begin{acknowledgements}
We acknowledge financial support by the German Research Foundation (J.P. Ruge: WO 857/10-1; H.H. Klahr: KL 14699-1). Computations were partially performed on the Bluegene/P supercomputer of MPG and the THEO cluster of MPIA at the Rechenzentrum Garching (RZG) of the Max Planck Society. We acknowledge the support of the International Max Planck Research School (IMPRS) of Heidelberg. The authors would like to acknowledge the helpful suggestions of the anonymous referee and the language editing by Amy Mednick.
\end{acknowledgements}
\vspace{-0.9cm}
\bibliography{bib}

\begin{thebibliography}{19}
\expandafter\ifx\csname natexlab\endcsname\relax\def\natexlab#1{#1}\fi

\bibitem[{{Cossins} {et~al.}(2010){Cossins}, {Lodato}, \&
  {Testi}}]{2010MNRAS.407..181C}
{Cossins}, P., {Lodato}, G., \& {Testi}, L. 2010, \mnras, 407, 181

\bibitem[{{Dohnanyi}(1969)}]{1969JGR....74.2531D}
{Dohnanyi}, J.~S. 1969, \jgr, 74, 2531

\bibitem[{{Gonzalez} {et~al.}(2012){Gonzalez}, {Pinte}, {Maddison},
  {M{\'e}nard}, \& {Fouchet}}]{2012A&A...547A..58G}
{Gonzalez}, J.-F., {Pinte}, C., {Maddison}, S.~T., {M{\'e}nard}, F., \&
  {Fouchet}, L. 2012, \aap, 547, A58

\bibitem[{{Isella} {et~al.}(2013){Isella}, {P{\'e}rez}, {Carpenter}, {Ricci},
  {Andrews}, \& {Rosenfeld}}]{2013ApJ...775...30I}
{Isella}, A., {P{\'e}rez}, L.~M., {Carpenter}, J.~M., {et~al.} 2013, \apj, 775,
  30

\bibitem[{{Jang-Condell}(2009)}]{2009ApJ...700..820J}
{Jang-Condell}, H. 2009, \apj, 700, 820

\bibitem[{{Papaloizou} \& {Lin}(1984)}]{1984ApJ...285..818P}
{Papaloizou}, J. \& {Lin}, D.~N.~C. 1984, \apj, 285, 818

\bibitem[{{P{\'e}rez} {et~al.}(2014){P{\'e}rez}, {Isella}, {Carpenter}, \&
  {Chandler}}]{2014ApJ...783L..13P}
{P{\'e}rez}, L.~M., {Isella}, A., {Carpenter}, J.~M., \& {Chandler}, C.~J.
  2014, \apjl, 783, L13

\bibitem[{{Quanz} {et~al.}(2013{\natexlab{a}}){Quanz}, {Amara}, {Meyer},
  {Kenworthy}, {Kasper}, \& {Girard}}]{2013ApJ...766L...1Q}
{Quanz}, S.~P., {Amara}, A., {Meyer}, M.~R., {et~al.} 2013{\natexlab{a}},
  \apjl, 766, L1

\bibitem[{{Quanz} {et~al.}(2013{\natexlab{b}}){Quanz}, {Avenhaus}, {Buenzli},
  {Garufi}, {Schmid}, \& {Wolf}}]{2013ApJ...766L...2Q}
{Quanz}, S.~P., {Avenhaus}, H., {Buenzli}, E., {et~al.} 2013{\natexlab{b}},
  \apjl, 766, L2

\bibitem[{{Ruge} {et~al.}(2013){Ruge}, {Wolf}, {Uribe}, \&
  {Klahr}}]{2013A&A...549A..97R}
{Ruge}, J.~P., {Wolf}, S., {Uribe}, A.~L., \& {Klahr}, H.~H. 2013, \aap, 549,
  A97

\bibitem[{{Sauter} \& {Wolf}(2011)}]{2011A&A...527A..27S}
{Sauter}, J. \& {Wolf}, S. 2011, \aap, 527, A27

\bibitem[{{van der Marel} {et~al.}(2013){van der Marel}, {van Dishoeck},
  {Bruderer}, {Birnstiel}, {Pinilla}, {Dullemond}, {van Kempen}, {Schmalzl},
  {Brown}, {Herczeg}, {Mathews}, \& {Geers}}]{2013Sci...340.1199V}
{van der Marel}, N., {van Dishoeck}, E.~F., {Bruderer}, S., {et~al.} 2013,
  Science, 340, 1199

\bibitem[{{Varni{\`e}re} {et~al.}(2006){Varni{\`e}re}, {Bjorkman}, {Frank},
  {Quillen}, {Carciofi}, {Whitney}, \& {Wood}}]{2006ApJ...637L.125V}
{Varni{\`e}re}, P., {Bjorkman}, J.~E., {Frank}, A., {et~al.} 2006, \apjl, 637,
  L125

\bibitem[{{Weingartner} \& {Draine}(2001)}]{2001ApJ...548..296W}
{Weingartner}, J.~C. \& {Draine}, B.~T. 2001, \apj, 548, 296

\bibitem[{{Wolf}(2003)}]{2003CoPhC.150...99W}
{Wolf}, S. 2003, Computer Physics Communications, 150, 99

\bibitem[{{Wolf}(2008)}]{2008PhST..130a4025W}
{Wolf}, S. 2008, Physica Scripta Volume T, 130, 014025

\bibitem[{{Wolf} \& {D'Angelo}(2005)}]{2005ApJ...619.1114W}
{Wolf}, S. \& {D'Angelo}, G. 2005, \apj, 619, 1114

\bibitem[{{Wolf} {et~al.}(1999){Wolf}, {Henning}, \&
  {Stecklum}}]{1999A&A...349..839W}
{Wolf}, S., {Henning}, T., \& {Stecklum}, B. 1999, \aap, 349, 839

\bibitem[{{Wolf} \& {Klahr}(2002)}]{2002ApJ...578L..79W}
{Wolf}, S. \& {Klahr}, H. 2002, \apjl, 578, L79

\end{thebibliography}

\end{document}